\documentclass[a4paper,11pt]{article}
\usepackage{jheppub}
\usepackage[T1]{fontenc}
\usepackage{float}
\usepackage{graphicx}  
\usepackage{dcolumn}   
\usepackage{color}
\usepackage{enumitem}
\usepackage[english]{babel} 
\usepackage{psfrag}
\usepackage{tikz}
\usepackage{tabularx}
\usepackage{bbm,bm}
\usepackage{appendix}
\numberwithin{equation}{section}
\usepackage{rotating}
\usepackage{epsfig}
\def\thesection{\arabic{section}}
\def\thesubsection{\arabic{section}.\arabic{subsection}}
\def\thesubsubsection{\arabic{section}.\arabic{subsection}.\arabic{subsubsection}}
\title{ Notes on exact solvability for rotating and pulsating strings in nonrelativistic Lifshitz background}
\author{ Adrita Chakraborty$^1$,}
 \affiliation{$^1$ Faculty of Physics and Applied Computer Science,\\ AGH University of Krakow, al. A. Mickiewicza 30, 30-059 Krakow, Poland}
\emailAdd{achakraborty@agh.edu.pl}
\abstract{
We construct one dimensional exactly solvable model by choosing a probe fundamental string rotating and pulsating in the planar Lifshitz spacetime that follows nonrelativistic Lifshitz scaling. We present suitable sets of embedding coordinates for rotating and pulsating strings to embed the string worldsheet on a hyperboloid with anisotropy-dependent eccentricity. The resulting worldsheet Lagrangians straightforwardly acquire the form of the Lagrangian of a Neumann-Rosochatius-type integrable model. Although the model assumes exact solutions for both the chosen ansatz its classical Liouville integrability is found to be conditional due to the presence of finite anisotropy in the target space geometry. We further use the exact solutions of the model to yield energy-momentum dispersion relations. We interpret those from the perspective of highly degenerate frustrated $J_1-J_2$ spin chain for rotating string and frustration-free Motzkin spin chain for pulsating string.}


\begin{document}
\flushbottom
\maketitle
\section{Introduction}
\label{section 1}
Unravelling the exact correspondence between gauge and gravity theories beyond the large $N$-limit is one of the prime and tantalizing areas of holographic research that are yet to explore. This requires the whole operator spectrum in the field theory and the associated energy spectrum in the gravity side.  Integrability provides a physical system with an exactly solvable structure due to larger underlying symmetries. These hidden symmetries appear as a closed algebra of an infinite tower of conserved quantities, also known as the integrals of motion. Previously, integrability was acquired on both sides of the famous AdS/CFT correspondence in the large N limit\cite{Maldacena:1997re, Gubser:1998bc, Witten:1998qj}.
It was established due to an exact match between the one-loop dilatation operators of the SU(2) sector of the$\mathcal{N}=4$ Super Yang-Mills theory with the Hamiltonian of the integrable SO(2) spin chain model\cite{Minahan:2002ve,Zarembo:2010sg,Babichenko:2009dk}. Recently, integrability has been further extended to extract operator spectrum beyond the large $N$ limit in CFTs of different dimensions by exploiting the exactly solvable features; see \cite{Levkovich_Maslyuk_2020} for a review. 
Added to this, the presence of at least approximate integrability has also been observed in various nonconformal holographic theories, for instance, large $N$ confining gauge theories\cite{Cooper:2014noa,Dubovsky:2015zey,Berenstein:2023lgo}
and nonrelativistic theories with Schrodinger symmetries\cite{Kawaguchi:2011wt,Guica:2017mtd}. 
\par The search for the dual string states of different gauge invariant operators in CFT gives rise to some generic class of semi-classical string ansatz. Some of these include circular rigid  rotating strings 
\cite{Minahan:2002rc} 
and pulsating strings \cite{Gubser:2002tv, Khan:2005fc,Khan:2003sm,Smedback:2004udl,Kruczenski:2004wg}. Rotating strings on sphere may further be classified into giant magnon\cite{Hofman:2006xt, Kruczenski:2004wg} 
and spiky strings\cite{Ishizeki:2007we,Mosaffa:2007ty,Hayashi:2007bq} 
which are eventually dual to elementary excitations with large momentum and higher twist operators respectively in the dual field theory description. On the other hand, pulsating or oscillating strings are more stable\cite{KHAN_2006} and dual to highly excited sigma model operators\cite{Gubser_2002}. In \cite{Kruczenski:2006pk, Arutyunov:2003uj, Arutyunov:2003za},
authors initiated a general prescription of reducing a classically integrable two-dimensional string sigma model to a one-dimensional Neumann-Rosochatius integrable model by implementing a generalized string ansatz. The latter is an extension of the famous classical Neumann system\cite{Neumann+1859+46+63}
and describes $N$ number of coupled harmonic oscillators moving in a unit sphere under the influence of an additional centrifugal inverse square potential\cite{RosochatiusAA}.
The solutions of this integrable model are capable of reproducing the finite gap energy spectra of a large class of generic string solutions\cite{Kazakov:2004qf, Beisert:2004ag}. Mechanical systems like Neumann, Rosochatius and Neumann-Rosochatius models play a crucial role to handle the problem of geodesics on ellipsoid or equivariant harmonic maps into sphere  \cite{Uhlenbeck,Babelon:2003qtg, Avan:1991ib,Avan:1989dn}. Plenty of research have been done so far\cite{Ahn:2008hj,Hernandez:2014eta,Hernandez:2015nba,Arutyunov:2016ysi,Hernandez:2017raj,Hernandez:2018gcd,Hernandez:2018lvh, Nieto:2018jzi, Chakraborty:2019gmt,Hernandez:2020igz, Chakraborty:2020las,Chakraborty:2022iuk,Chakraborty:2022eeq,Hernandez:2022dmf} 
to explore general rotating and pulsating string solutions in different choices of 10D AdS/S compact spaces with pure or mixed flux from the systematic construction of the one-dimensional Lagrangian of NR system.


\par  A well-known class of solutions of Einstein-Proca type gravity theories follows nonrelativistic Lifshitz symmetry defined as
\begin{equation}
    t\rightarrow \lambda^z t,~~x_i\rightarrow \lambda x_i
\end{equation}These are ddesignated as some special forms of the Schrodinger geometries. However, it has been found to be chaotic due to the finite anisotropy index in the temporal scaling. Previously, different semiclassical string ansatz in Lifshitz spacetime resulted in chaotic string solutions\cite{Giataganas:2014hma,Bai:2014wpa}. In this article we aim to reproduce a 1d classical Neumann-Rosochatius-type exactly solvable model from the 2d string sigma model constructed for a probe fundamental string moving in a nonrelativistic Lifshitz spacetime that follows anisotropic Lifshitz scaling. We start with the Lif$_3\times H_2$ background, a typical solution of type IIB supergravity equations of motion. . Here $H_2$ represents a 2D hyperboloid which is an internal 2D space with constant curvature that may be compactified. These solutions are obtained in \cite{Gregory_2010}
via compactifications of $\mathcal{N}$ = 4 5D
gauged supergravity, which can be uplifted to Type IIB supergravity in ten dimensions, or
eleven-dimensional supergravity. We postulate a suitable set of generic Neumann-Rosochatius ansatz of both rotating and pulsating strings for the embedding of the string worldsheet into the 2+1D Lifshitz spacetime. In such ansatz, the radial coordinate $r$ appears as a function of the anisotropy index $z$. Both rotating and pulsating strings are considered to contain momenta along any of the spatial directions of the Lif$_3$ and the directions along $H_2$ are localized during our study. Very recently, renormalizable $\mathcal{N}=1$ supersymmetric sigma models with anisotropic Lifshitz symmetry have been constructed and merged to the relativistic sigma model at low energy limit\cite{Yan_2023}.
Here we formulate a 2D classical string sigma model starting with our chosen string ansatz in the anisotropic Lifshitz background. We show that, for any arbitrary anisotropy index $z$, both of these systems become equivalent to an exactly solvable classical model containing $r^2$ and $\frac{1}{r^2}$-type potentials similar to the integrable Neumann-Rosochatius model for constant Lagrange multiplier. However, involution of the constants of motion appears only when $z$ assumes a specific functional form involving other different parameters present in the system. Furthermore, the equations of motion of the model constructed herein are employed to study the scaling between conserved energy and momenta for both of the string ansatz. Similarly as in AdS/CFT duality, we assert a possible interpretation of these scaling relations in terms of spin chains that are already observed to persist integrability at certain conditions and are related to the Lifshitz field theories dual to our chosen background through the well-known Lifshitz holography\cite{Taylor_2016}. We speculate that the frustrated $J_1-J_2$ spin chain with massive ground state degeneracy (see for review, 7th chapter of \cite{Pavarini:205123}, \cite{Balents_2016}) and the frustration-free Motzkin spin chains \cite{Movassagh_2017} in the continuum limit can be prescribed as the plausible equivalent spin chain descriptions respectively for the rotating and pulsating strings developed from the exactly solvable framework in our article.

In section \ref{section 2},  we present the construction of deformed Lagrangian, Hamiltonian and the corresponding Uhlenbeck integrals of motion of exactly solvable Neumann-Rosochatius-type model for closed rigid rotating and pulsating strings, both having momenta along one of the spatial directions of 2+1D $\text{Lif}_{3}$ background. We derive in section \ref{section 3} harmonic oscillator-type solutions for rotating string and oscillation number-dependent solutions for pulsating string by solving the solvable equations of motion of the classical models constructed thereby. Using these solutions, we further study the scaling relations between the conserved charges and energy-momenta dispersion relations for both the closed rotating and pulsating string states in subsections \ref{subsection 3.1} and \ref{subsection 3.2} respectively. Section \ref{section 4} includes some plausible comments on the suitable dual spin chain descriptions for different string states. Finally, we conclude our results in section \ref{section 5} with a summary of our present work and its future prospects.
\section{ Neumann-Rosochatius model from 2+1D Lifshitz target spacetime}
\label{section 2}
Despite being a nonrelativistic gravity background with Lorentz breaking Lifshitz symmetry, the 2D  worldsheet Lagrangians for fundamental rotating and pulsating strings in Lifshitz spacetime is shown in this section to emerge as the Lagrangian of a 1D classical Neumann-Rosochatius-like model. The 2+1D Lifshitz solutions $\text{Lif}_{3}\times H_2$ of type IIB supergravity equations of motion is given by
\begin{equation}
ds^{2}=G_{MN}d x^{M}dx^{N}=L^2\left(r^{2z}dt^2-r^2dx^2-\frac{dr^2}{r^2}\right)+dH_2^2\ .
\label{Lifshitz metric}
\end{equation}Note that, the metric (\ref{Lifshitz metric}) breaks the conformal supersymmetry. This is rendered by the appearance of an extra parameter $z$ which assumes the values such that $z\geq 1$ to be compatible with the scenario of Lifshitz holography. This parameter defines the Lifshitz fixed point and is known as the dynamical critical exponent. It further introduces anisotropy between the temporal and spatial directions of the background. We wish to probe the above gravity background with fundamental bosonic string whose Polyakov action with conformal gauge can be written as 
\begin{eqnarray}
S=\frac{T}{2}\int d\tau d\sigma \sqrt{-\gamma}
\gamma^{\alpha\beta}G_{\alpha\beta}-\frac{T}{2}
\int d\sigma d\tau \epsilon^{\alpha\beta}B_{MN}\partial_\alpha X^M\partial_\beta X^N \ ,
\label{polyakov}
\end{eqnarray}
where $G_{\alpha\beta}=G_{MN}\partial_\alpha X^M\partial_\beta X^N$, $X^N(\tau,\sigma), M, N =0,\dots,9$ are the string embedding
coordinates. Further, $\partial_\alpha\equiv \frac{\partial}{\partial \sigma^\alpha}\ ,~  \sigma^0=\tau,~ \sigma^1=\sigma$ and $T$ is the string
tension. The string is usually embedded into a generic ten dimensional background with the metric $G_{MN}$  and NS-NS two form field $B_{MN}$. Finally
$\gamma_{\alpha\beta}$ is two dimensional world-sheet metric whose equations of motion gives the stress-energy tensor as
\begin{equation}\label{eqgamma}
T_{\alpha\beta}=-\frac{2}{\sqrt{-\gamma}}\frac{\delta S}{\delta \gamma^{\alpha\beta}}=
-\frac{T}{2}\gamma_{\alpha\beta}\gamma^{\gamma\delta}G_{\gamma\delta}+TG_{\alpha\beta}=0 \ . 
\end{equation}
\subsection{Embedding coordinates and constraints }
\label{subsection 2.1}
As anti de-Sitter spacetime is a vacuum solution of Einstein's field equations with constant negative curvature it is customary to embed d-dimensional anti de-Sitter spacetime on a global d+1-dimensional hyperboloid constrained with
\begin{equation}
    -X_0^2+X_1^2+...+X_{d-1}^2-X_d^2=-1\,.
    \label{constraint 1}
\end{equation}Though the 2+1D Lifshitz spacetime is a Lorentz-breaking deformation of the usual AdS it also has a negative curvature $R=-2(z^2+z+1)$ that depends on anisotropy parameter $z$. Thus for any fixed value of $z$, the Lifshitz spacetime assumes constant negative curvature. This makes it possible to consider Lifshitz geometry to be a non-compact one and consistent to be embedded on a negatively curved hyperboloid space with an extra dimension. The metric given in (\ref{Lifshitz metric}), has a form analogous to the Poincare patch of AdS global metric. We embed the string worldsheet in our chosen Lifshitz spacetime on a 3+1D hyperboloid. The embeddings are chosen as a general ansatz which reproduces a large class of rotating as well as pulsating string states moving in the background (\ref{Lifshitz metric}) when uplifted to its 10D counterpart. Let us write the embeddings as
\begin{equation}
    X_0+iX_3=r_1e^{i\Phi_1},~~X_1+iX_2=r_2e^{i\Phi_2}\,,
\end{equation}where, $X_i$ are the suitably chosen coordinates of the higher dimensional hyperboloid. This gives from (\ref{constraint 1})
\begin{equation}
    r_1^2-r_2^2=1
\end{equation}which is the equation of a hyperbola with equal semi-major and semi-minor axes. This embedding gives the metric (\ref{Lifshitz metric}) from the relation 
\begin{equation}
    ds^2=-|dX_0|^2+|dX_1|^2+|dX_2|^2-|dX_3|^2
    \label{metric}
\end{equation}Let us construct these embeddings for probe strings in 2+1D Lifshitz spacetime by choosing  
\begin{equation}
  \Phi_1\equiv t,~~  \Phi_2\equiv x,~~r_2\equiv r^z,~~r_1\equiv r\,.
    \label{embedding}
\end{equation}Comparing (\ref{metric}) with (\ref{Lifshitz metric}) yields the relations 
\begin{equation}
    r_2=r_1^z,~~r_1^2-z^2r_2^2=1
    \label{relation 1}
\end{equation}It is evident that the constraint is now modified into the 2nd relation in (\ref{relation 1}) which describes a hyperbola with unit semimajor axis while the semiminor axis has a length $\frac{1}{z^2}$. This deformation occurs in the hyperbolic geometry because of the anisotropy between the temporal and spatial directions in the chosen Lifshitz-type supergravity solution. Therefore, the anisotropy in the background introduces modified constraints while embedding it on a higher dimensional hyperboloid. At this stage we will consider the dynamics of the probe fundamental string only along the direction of $2+1$D Lifshitz spacetime and hence we may localize it at some constant point along the directions of the 2D hyperboloid space $H_2$.

\subsection{ Lagrangian of Neumann-Rosochatius model and integrals of motion}
\label{subsection 2.2}
 Here we wish to check whether the nonrelativistic Lifshitz system probed by fundamental string can be reduced into an integrable Neumann-Rosochatius model. To verify this, we must reduce the target space Lagrangian into a 2D worldsheet Lagrangian of the string moving inside the target space background. This requires a choice of some typical string ansatz. We take the ansatz for rotating and pulsating string solutions respectively as 
 \begin{equation}
    t=\phi_1(\tau,\sigma)=\kappa\tau,~~ r_1=r_1(\sigma),~~ r_2=r_2(\sigma),~~x=\phi_2(\tau,\sigma)=\omega\tau+f(\sigma)
    \label{rotating}
 \end{equation}and
 \begin{equation}
      t=\phi_1(\tau,\sigma)=\tau,~~ r_1=r_1(\tau),~~ r_2=r_2(\tau),~~x=\phi_2(\tau,\sigma)=m\sigma+f(\tau)
      \label{pulsating}
 \end{equation} $\tau$ and $\sigma$ represent the coordinates of the 1+1D worldsheet locally swept out by the probe string on the embedded hyperbolic Lifshitz geometry. Using  these local coordinates in the action (\ref{polyakov}) the worldsheet Lagrangian may be written as 
 \begin{equation}
 \begin{split}
 \mathcal{L} = \frac{T}{2}\left[r_2^2\Big\{\left(\partial_{\sigma}\phi_1\right)^2-\left(\partial_{\tau}\phi_1\right)^2\Big\}+r_1^2\Big\{\left(\partial_{\sigma}\phi_2\right)^2-\left(\partial_{\tau}\phi_2\right)^2\Big\}\right. \\ \left.+\left(1-\frac{z^2r_2^2}{r_1^2}\right)\Big\{\left(\partial_{\sigma}r_1\right)^2-\left(\partial_{\tau}r_1\right)^2\Big\}\right]+\frac{\Lambda_1}{2}\left(r_1^2-z^2r_2^2-1\right)
   \end{split}
 \label{lagrangian 1}
   \end{equation}
   Here $\Lambda$ and $\Lambda_{1}$ are suitably chosen Lagrange multipliers corresponding to the constraints. For rotating string ansatz (\ref{rotating}), this Lagrangian reduces to 
   \begin{equation}
       \mathcal{L}=\frac{T}{2}\left[\kappa^2r_2^2+(f^{'2}-\omega^2)r_1^2+\left(1-\frac{z^2r_2^2}{r_1^2}\right)r_1^{'2}\right]+\frac{\Lambda_1}{2}\left(r_1^2-z^2r_2^2-1\right)
       \label{Lagrangian 2}
   \end{equation}Now, using the relations in (\ref{relation 1}), we may write $\left(1-\frac{z^2r_2^2}{r_1^2}\right)r_1^{'2}=r_1^{'2}-r_2^{'2}$ so that the Lagrangian (\ref{Lagrangian 2}) becomes
   \begin{equation}
         \mathcal{L}=\frac{T}{2}\left[r_1^{'2}-r_2^{'2}+\kappa^2r_2^2+(f^{'2}-\omega^2)r_1^2\right]+\frac{\Lambda_1}{2}\left(r_1^2-z^2r_2^2-1\right)
   \end{equation}Similarly the pulsating string ansatz (\ref{pulsating}) yields the Lagrangian
   \begin{equation}
           \mathcal{L}=\frac{T}{2}\left[\dot{r}_1^{2}-\dot{r}_2^{2}+r_2^2+(m^2-\dot{f}^{2})r_1^2\right]+\frac{\Lambda_1}{2}\left(r_1^2-z^2r_2^2-1\right)
   \end{equation}The equation of motion for the angular coordinate $f$ gives $f^{'}=\frac{C_1}{r_1^2}$ for rotating string case and $\dot{f}=\frac{C_2}{r_1^2}$ for pulsating string case, $C_1$ and $C_2$ being some suitable integration constants which we will derive in the next section. Thus, substituting these in respective Lagrangians we respectively get for rotating and pulsating string ansatz,
   \begin{equation}
        \mathcal{L}=\frac{T}{2}\left[r_1^{'2}-r_2^{'2}+\kappa^2r_2^2+\frac{C_1^2}{r_1^2}-\omega^2r_1^2\right]+\frac{\Lambda_1}{2}\left(r_1^2-z^2r_2^2-1\right)
        \label{rotating Lagrangian}
   \end{equation}and 
   \begin{equation}
    \mathcal{L}=\frac{T}{2}\left[\dot{r}_1^{2}-\dot{r}_2^{2}+r_2^2+m^2r_1^2-\frac{C_2^2}{r_1^2}\right]+\frac{\Lambda_1}{2}\left(r_1^2-z^2r_2^2-1\right)
    \label{pulsating lagrangian}
    \end{equation}The equations of motion for $r_1$ and $r_2$ for the Lagrangian (\ref{rotating Lagrangian}) are
    \begin{equation}
r_1^{''}+\left(\omega^2-\frac{\Lambda_1 }{T}\right)r_1+\frac{C_1^2}{r_1^3}=0
\label{EOM1}
    \end{equation}and
    \begin{equation}
    r_2^{''}+(\kappa^2-\frac{\Lambda_1 z }{T})r_2=0  
    \label{EOM2}  \end{equation}These may be considered as two coupled equations of motion which can also be derived from the Lagrangian
    \begin{equation}
        L_{R}=r_1^{'2}-r_2^{'2}+\kappa^2r_2^2+\frac{C_1^2}{r_1^2}-\omega^2r_1^2+\frac{A_1}{2}\left(r_1^2-z^2r_2^2-1\right)
        \label{rotating NR}
    \end{equation}where $A_1=\frac{\Lambda_1}{T}$. Similarly, the equations of motion from the pulsating string Lagrangian (\ref{pulsating lagrangian}) are
    \begin{equation}
         \ddot{r}_1-m^2r_1+\frac{C_2^2}{r_1^3}-\frac{\Lambda_1}{T}r_2=0
        \label{EOM3}
    \end{equation}and
      \begin{equation}
   \ddot{r}_2-\left(1-\frac{\Lambda_1z}{T}\right)r_2=0
        \label{EOM4}
    \end{equation}These coupled equations of motion can again be the Euler-Lagrange equations of the Lagrangian
    \begin{equation}
        L_{P}=\dot{r}_1^{2}-\dot{r}_2^{2}+r_2^2+\frac{C_2^2}{r_1^2}-m^2r_1^2+\frac{A_1}{2}\left(r_1^2-z^2r_2^2-1\right)
        \label{pulsating NR}
    \end{equation}The Hamiltonians of these systems may be written in the forms as 
    \begin{equation}
           H_R=r_1^{'2}+r_2^{'2}-\kappa^2r_2^2+\omega^2r_1^2-\frac{C_1^2}{r_1^2}-\frac{A_1}{2}(r_1^2-z^2r_2^2-1)
           \label{rotating hamiltonian}
    \end{equation}and 
    \begin{equation}
           H_P=\dot{r}_1^{2}+\dot{r}_2^{2}-r_2^2+m^2r_1^2-\frac{C_2}{r_1^2}-\frac{A_1}{2}(r_1^2-z^2r_2^2-1)
           \label{pulsating hamiltonian}
    \end{equation}respectively. The Lagrangians and Hamiltonians found in (\ref{rotating NR}), (\ref{pulsating NR}), (\ref{rotating hamiltonian}) and (\ref{pulsating hamiltonian}) contain potentials with $r^2$ and $\frac{1}{r^2}$ forms which represents harmonic oscillator-type and centrifugal-type motion respectively. Hence, it is obvious that, when probed by rotating as well as pulsating string, the 2D string worldsheet Lagrangian even for nonrelativistic anisotropic Lifshitz background reduces to a one dimensional Lagrangian describing the motion of coupled harmonic oscillators constrained on a hyperbola with eccentricity $\pm\frac{\sqrt{1+z^2}}{z}$ in the presence of a centrifugal potential barrier. Such system can thus be thought to be similar as the integrable classical Neumann-Rosochatius model constrained on the hyperbolae of the above kinds. Previously it was shown that for some particular combinations of the anisotropy parameters, the nonrelativistic candidate theories follow classical Liouvillian integrability\cite{Giataganas:2014hma}. However, here it is shown that a suitably chosen rotating and pulsating string ansatz that embed the string worldsheet on a hyperbolic geometry with Lifshitz scaling can reduce the system into an exactly solvable model similar to the classically integrable Neumann-Rosochatius one for any arbitrary value of $z$.
\subsection{Integrals of motion}
    The typical form of the constants of motion in involution for the Liouville integrable Neumann-Rosochatius model was first developed by K.Uhlenbeck in\cite{10.1007/BFb0069763}.
    and it is expressed as 
    \begin{equation}
    I_{i}= x_{i}^{2}+\sum_{j\neq i}\frac{1}{\omega_{i}^{2}-\omega_{j}^{2}}\left[\left(x_{i}x_{j}^{'}-x_{j}x_{i}^{'}\right)^{2}+\pi_{\alpha_ i}^{2}\frac{x_{j}^{2}}{x_{i}^{2}}+\pi_{\alpha_ j}^{2}\frac{x_{i}^{2}}{x_{j}^{2}}\right], \
\label{Integrals}
    \end{equation}Here, $x_i$'s are general position coordinates and $\alpha_i$'s are the angular coordinates, $\pi_{\alpha_i}$'s are the canonically conjugate momenta for the angular coordinates and $\omega_i$'s are the angular frequencies of the coupled harmonic oscillators in the Neumann-Rosochatius model. For our choice of ansatz given in (\ref{rotating}) and (\ref{pulsating}), the angular frequencies are $\kappa$ and $\omega$ as $t$ and $x$ are assumed to have angular interpretations. The angular coordinates of the NR model constructed herein are denoted by $f(\sigma)$ whose corresponding canonically conjugate momenta for rotating and pulsating string cases are derived as $(\pi_{f})_R=C_1$ and $(\pi_f)_P=C_2$ respectively. Hence the Uhlenbeck constants of motion for the NR Lagrangians (\ref{rotating NR}) and (\ref{pulsating lagrangian}) are given as 
    \begin{equation}
         I_{1}= r_{1}^{2}+\frac{1}{\kappa^{2}-\omega^{2}}\left[\left(r_{1}r_{2}^{'}-r_{2}r_{1}^{'}\right)^{2}+\frac{C_1^2r_{2}^{2}}{r_{1}^{2}}\right], \
    \end{equation}and 
    \begin{equation}
          I_{1}= r_{1}^{2}+\frac{1}{m^{2}}\left[\left(r_{1}\dot{r}_{2}-r_{2}\dot{r}_{1}\right)^{2}+\frac{C_2^2r_{2}^{2}}{r_{1}^{2}}\right]
    \end{equation}For an NR model constrained on any hyperbolic geometry and having two Uhlenbeck integrals of motion, the condition, $I_1-I_2=1$ is supposed to be satisfied. Thus, for both cases, the other integral of motion can be obtained from the constraint $I_1-I_2=1$. Now, the modified constraint in (\ref{relation 1}) is supposed to introduce nontrivial deformations in the expressions of these Uhlenbeck integrals of motion. Those deformations can be expected to be the functions of $r_1, r_2$ and $z$. Let us denote them as $u(r_1,r_2,z)$ and $v(r_1,r_2,z)$ respectively for rotating and pulsating string cases. With these deformations, let us write the integrals of motion for rotating and pulsating string cases in a convenient form as 
    \begin{equation}
        I_{1}= r_{1}^{2}+\frac{1}{\kappa^{2}-\omega^{2}}\left[\left(r_{1}r_{2}^{'}-r_{2}r_{1}^{'}\right)^{2}+\frac{C_1^2r_{2}^{2}}{r_{1}^{2}}+2u(r_1,r_2,z)\right], \
        \label{deformed IOM1}
        \end{equation}and 
        \begin{equation}
            \tilde{I}_{1}= r_{1}^{2}+\frac{1}{m^{2}}\left[\left(r_{1}\dot{r}_{2}-r_{2}\dot{r}_{1}\right)^{2}+\frac{C_2^2r_{2}^{2}}{r_{1}^{2}}+2v(r_1,r_2,z)\right], \label{deformed IOM 2}
        \end{equation}respectively. To identify the exact functional forms of these deformations, we will use the condition, $I_1^{'}=0$. We will use a constant Lagrange multiplier throughout the entire analysis from now on. Using the equations of motion for $r_1$, $r_2$ and the constraint $r_1^2-z^2r_2^2=1$, we get the expression for the deformation $u$ from $I_1^{'}=0$ as
        \begin{equation}
            u=-\left[\frac{\kappa^2-\omega^2}{2}-\frac{1}{2}\left(\frac{\Lambda_1}{T}-\frac{z\Lambda_1}{T}-\kappa^2-\omega^2\right)\right]r_1^2-\frac{C_1^2r_2^2}{r_1^2}
        \end{equation}Substituting for $u$ in (\ref{deformed IOM1}), we get the integrals of motion 
        \begin{equation}
            I_1=\left(1-2\kappa^2+\frac{\Lambda_1}{T}-\frac{\Lambda_1 z}{T}\right)r_1^2+\frac{1}{\kappa^2-\omega^2}\left[\left(r_2r_1^{'}-r_1r_2^{'}\right)^2-\frac{C_1^2r_2^2}{r_1^2}\right]
        \end{equation} If we consider the other Uhlenbeck integral of motion as 
        \begin{equation}
            I_2=r_{2}^{2}+\frac{1}{\kappa^{2}-\omega^{2}}\left[\left(r_{1}r_{2}^{'}-r_{2}r_{1}^{'}\right)^{2}+\frac{C_1^2r_{2}^{2}}{r_{1}^{2}}+2\tilde{u}\right]
        \end{equation}with $\tilde{u}$ as the relevant deformation, then using the constraint $r_1^2-z^2r_2^2=1$ and $I_1-I_2=1$ and substituting the expression similarly derived for $\tilde{u}$, we can write $I_2$ as
        \begin{equation}
            I_2=\left(1-2\kappa^2+\frac{\Lambda_1}{T}-\frac{\Lambda_1 z}{T}\right)z^2r_2^2+\left(-2\kappa^2+\frac{\Lambda_1}{T}-\frac{\Lambda_1 z}{T}\right)+\frac{1}{\kappa^2-\omega^2}\left[\left(r_2r_1^{'}-r_1r_2^{'}\right)^2-\frac{C_1^2r_2^2}{r_1^2}\right]
        \end{equation}For any integrable system, we require all the constants of motion to be in involution, i.e., $\Big\{I_i,I_j\Big\}=0$. Although, the integrals of motion $I_1$ and $I_2$ commutes with the canonical momentum $\pi_f$, they commute with each other only when $z$ assumes a definite functionall form of the string tension $T$, Lagrange multiplier $\Lambda_1$ and $\kappa$ as
        \begin{equation}
            z=1-\frac{T}{\Lambda_1}\left(1-2\kappa^2\right).
            \label{integrability}
        \end{equation} Therefore, the integrable feature of the equivalent Neumann-Rosochatius-type model for rotating string ansatz in anisotropic nonrelativistic Lifshitz spacetime is conditional and it occurs due to the presence of finite anisotropy parameter $z$ in the target space. We can find the deformation $v$ in the expression $\tilde{I}_1$ for pulsating string ansatz. This will give
        \begin{equation}
            v=-\left[\frac{m^2}{2}-\frac{1}{2}\left(\frac{\Lambda_1}{T}-\frac{z\Lambda_1}{T}-m^2\right)\right]r_1^2-\frac{C_1^2r_2^2}{r_1^2}
        \end{equation}This gives
        \begin{equation}
            \tilde{I}_1=\left(1-2m^2+\frac{\Lambda_1}{T}-\frac{z\Lambda_1}{T}\right)r_1^2+\frac{1}{m^{2}}\left[\left(r_{1}\dot{r}_{2}-r_{2}\dot{r}_{1}\right)^{2}-\frac{C_2^2r_{2}^{2}}{r_{1}^{2}}\right]
        \end{equation}Then, using the geometrical constraint $r_1^2-z^2r_2^2=1$ and the constraint $\tilde{I}_1-\tilde{I}_2=1$ for the integrals of motion, we get the other integral of motion as 
        \begin{equation}
            \tilde{I}_2=\left(1-2m^2+\frac{\Lambda_1}{T}-\frac{z\Lambda_1}{T}\right)z^2r_2^2+\left(\frac{\Lambda_1}{T}-\frac{z\Lambda_1}{T}-2m^2\right)+\frac{1}{m^{2}}\left[\left(r_{1}\dot{r}_{2}-r_{2}\dot{r}_{1}\right)^{2}-\frac{C_2^2r_{2}^{2}}{r_{1}^{2}}\right]
        \end{equation}Similarly as the case of rotating string, the integrability of Neumann-Rosochatius-type model for pulsating string also shows when the condition 
        \begin{equation}
            z=1-\frac{T}{\Lambda_1}(1-2m^2)
        \end{equation}is satisfied by the anisotropy parameter $z$.
   

\section{Exact solutions and energy dispersion relations }
\label{section 3}
We will solve the equations of motion (\ref{EOM1}), (\ref{EOM2}) for rotating string  and (\ref{EOM3}), (\ref{EOM4}) for pulsating string to understand the string profiles as solutions of the associated equivalent solvable models. The conserved quantities corresponding to the cyclic coordinates $t$ and $\phi$ in the Lagrangian (\ref{lagrangian 1}) are given as energy 
\begin{equation}
    E=-\int_{0}^{2\pi}d\sigma\frac{\partial\mathcal{L}}{\partial(\partial_{\tau}t)}
    \label{energy}
\end{equation}and angular momentum \begin{equation}
    J=\int_{0}^{2\pi}d\sigma\frac{\partial\mathcal{L}}{\partial(\partial_{\tau}x)}\,.
    \label{angular momentum}
\end{equation}The solutions of the equations of motion of the model formulated in subsection \ref{subsection 2.2} are expected to help us in computing the energy-angular momenta scaling dispersion relations for both rotating and pulsating string ansatz.
\subsection{Rotating string with one angular momentum}
\label{subsection 3.1}
\begin{figure}
\includegraphics[scale=0.38]{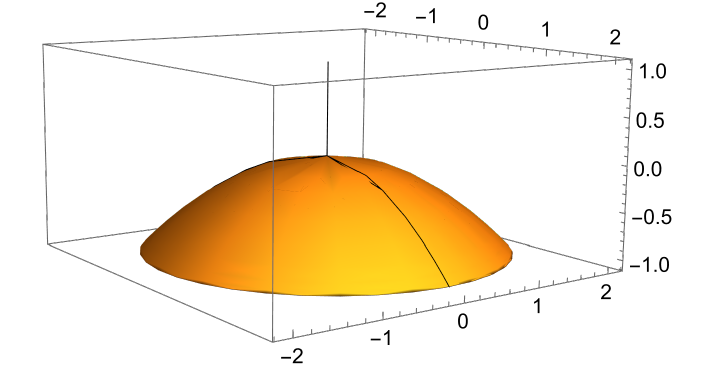}
\hfill
\includegraphics[scale=0.38]{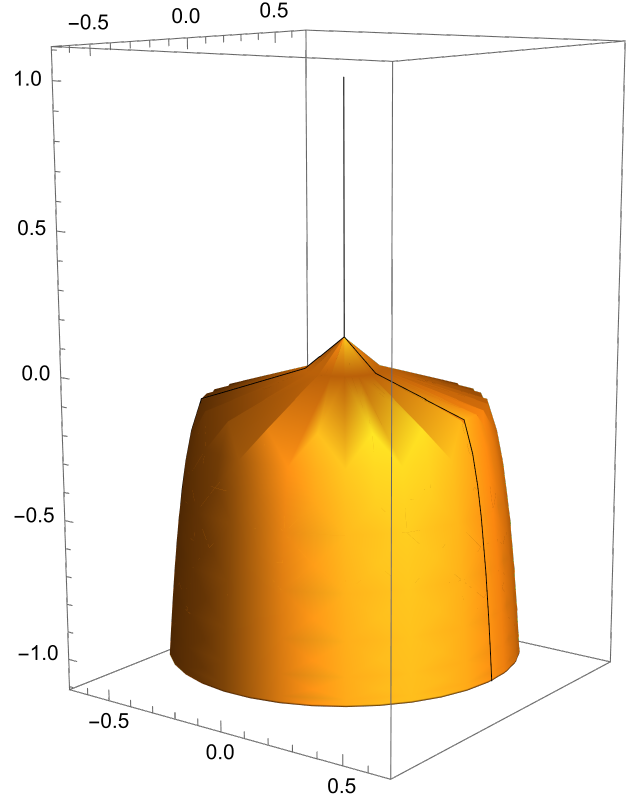}
\hfill
\includegraphics[scale=0.38]{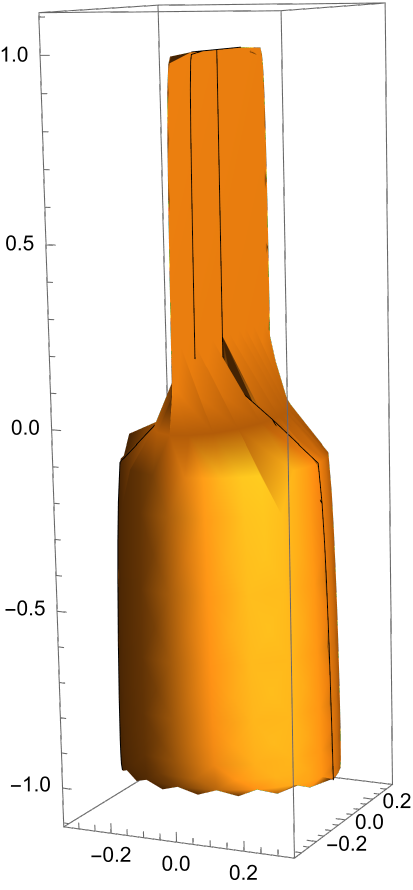}
\caption{String embeddings in (\ref{rotating 2}) for $z=2$, $z=10$ and $z=20$ (Left to right). Here, we fix $n=1$ and $\kappa=1$. The values of the constant $A_1$ varies accordingly with the relation $\kappa^2-A_1z=n$, for each of the chosen $z$ values.}
\label{fig:1}
\end{figure}

The equation of motion for $r_2(\sigma)$ for the rotating string may be written as
\begin{equation}
    r_2^{''}(\sigma)=-kr_2
    \label{EOM5}
\end{equation}where $k=\kappa^2-\frac{\Lambda_1 z}{T}=\kappa^2-A_1z$. This equation clearly describes the motion of a harmonic oscillator with frequency $k$. This has a straightforward solution as $r_1(\sigma)=pe^{ik\sigma}+qe^{-ik\sigma}$, $p$ and $q$ are some suitable constants which may be derived by using some proper initial conditions. When we consider a closed rigid rotating string then the coordinate $r_i(\sigma)$'s follow the conditions 
\begin{equation}
    r_i(\sigma)|_{\sigma=0}=r_i(\sigma)|_{\sigma=2\pi}
    \label{initial condition}
\end{equation}This gives $p=q$ and we get $\cos{(2k\pi)}=1$ which further yields $2k\pi=2n\pi,~~n\in\mathbb{Z}_+$. This eventually yields $\kappa^2-A_1 z=n,~~n\in \mathbb{Z}_+$. Thus the solution of (\ref{EOM5}) becomes
\begin{equation}
    r_2(\sigma)=p\left(e^{in\sigma}
+e^{-in\sigma}\right),~~n\in\mathbb{Z}_+
\end{equation}Again, from the first relation of (\ref{relation 1}), we achieve 
\begin{equation}
    r_1(\sigma)=p^{\frac{1}{z}}\cos^{\frac{1}{z}}(2n\sigma)
\end{equation}and followed by (\ref{initial condition}) we can get $p=1$. Thus the complete solution can be written as
\begin{subequations}
    \begin{align}
         &r_1(\sigma)=\left(e^{in\sigma}
+e^{-in\sigma}\right)^{\frac{1}{z}}\\&\label{r1 solution}
  r_2(\sigma)=\left(e^{in\sigma}
+e^{-in\sigma}\right),~~n\in \mathbb{Z}_+
    \end{align}
\end{subequations}One can substitute these solutions into the embeddings given in (\ref{embedding}) and use  the rotating string ansatz (\ref{rotating}) to have 
\begin{equation}
    \tilde{X}=\sqrt{X_1^2+X_2^2}=\cos{(n\sigma)},~X_0=\sqrt{\cos{(n\sigma)}}\cos{(\kappa\tau)},~X_3=\sqrt{\cos{(n\sigma)}}\sin{(\kappa\tau)}
    \label{rotating 2}
\end{equation}We plot these embeddings in fig.(\ref{fig:1}) for different values of the anisotropy index $z$, provided that these values are selected with associated values of $\kappa$, $n$ and $A_1$ which satisfy the relation $\kappa^2-A_1z=n$. For smaller $z$ values, it shows rotation of strings which swipes out the geometry of an approximate hemisphere. When we start to increase the values of $z$, i.e., with larger anisotropy index the ansatz gives rotating strings swiping out a geometry that gradually becomes cylindrical.  Assuming that these solutions satisfy the equations of motion (\ref{EOM1}) and (\ref{EOM2}) we can derive the suitable forms of different constants that we included in our computation so far. While doing so, we obtain the expression of $r_1(\sigma)$ in (\ref{r1 solution}) as a consistent solution of equation (\ref{EOM1}) when we take $\omega^2=A_1$ and $\tan^2(n\sigma)=\frac{n^2}{1-z}$.
Note that, among the Virasoro constraints $G_{\tau\tau}+G_{\sigma\sigma}=0$ and $G_{\tau\sigma}=0$ achieved from the action (\ref{polyakov}), the second one gives $r^2\omega\frac{C_1}{r_1^2}=0$. This gives after substituting $r=r_1$, $\omega C_1=0$ and for nonzero angular frequency $C_1=0$. This provides the exactly solvable model for rotating string, with a structure similar to the Neumann integrable model. Such model represents the motion of a harmonic oscillator without the influence of any centrifugal inverse square potential. At this stage, we have from the expression of $\omega$ and $\kappa$
\begin{equation}
    \kappa=\pm \sqrt{\omega^2z+n}
\end{equation}
The conserved energy and angular momentum may be deduced from (\ref{energy}) and (\ref{angular momentum}) as
\begin{equation}
    E=-\frac{2T\kappa}{n}\left[2n\pi+\frac{\sin(8n\pi)}{4}\right]=-4T\kappa\pi
\end{equation}and
\begin{equation}
    J=-\frac{T\omega}{2n}\left(\cos^{\frac{2}{z}-1}x\sin x+(\frac{2}{z}-1)\cos^{\frac{2}{z}-3}x\sin x+....+\frac{\sin x}{4}+\frac{x}{2}\right)\Big\|_{0}^{4n\pi}
\end{equation}
From these expressions, we have the scaling as 
\begin{equation}
   E=\frac{2\kappa}{\omega}J=\pm\frac{2J}{\omega}\sqrt{\omega^2z+n}
   \label{scaling}
\end{equation}
Thus the energy of the rotating string states in Lifshitz background is scaled linearly with the angular momentum $J$ with a scaling factor that depends nontrivially on the critical exponent $z$. Fixing the values of $n$ and $\omega$, we may see the energy $E$ increases monotonically for higher values of $z$. 

\subsection{Pulsating string with one angular momentum}
\label{subsection 3.2}
Pulsating or oscillating string solutions are customarily expressed by using the large adiabatic oscillation number. We implement our ansatz (\ref{pulsating}) and (\ref{embedding}) along with different constraints (\ref{relation 1}) given in subsection \ref{subsection 2.1} in the Virasoro constraint $G_{\tau\tau}+G_{\sigma\sigma}=0$ and $G_{\tau\sigma}=0$. We choose the string pulsating with one angular momentum $J$ along the direction $r$ in the chosen Lifshitz background. The oscillation number can thus be written as 
\begin{equation}
    N=T\int \dot{r}dr=T\int \dot{r}_1 dr_1
\end{equation}Let us write $\mathcal{N}=\frac{N}{T}=\int \dot{r}_1 dr_1$. Again from Virasoro constraints we get
\begin{subequations}
    \begin{align}
        &r_2^2-C_2^2-r_1^2m^2-\left(1-\frac{z^2r_2^2}{r_1^2}\right)\dot{r}_1^2=0   \label{virasoro 1}\\&
        mC_2^2=0
        \label{virasoro 2}
    \end{align}
\end{subequations}The energy and angular momentum are given as 
\begin{equation}
        \mathcal{E}=\frac{E}{T}=r_2^2,~~\mathcal{J}=\frac{J}{T}=-C_2
\end{equation}It is thus obvious that a finite nonzero value of $C_2$ demands string winding number $m=0$ from (\ref{virasoro 2}). 
Hence we get from (\ref{virasoro 1}) 
\begin{equation}
    \dot{r}_1=r_1\sqrt{r_2^2-C_2^2}
\end{equation}where we used the relation $r_1^2-z^2r_2^2=1$. Again, using $r_2=r_1^z$, we have $r_1^2=\mathcal{E}^{\frac{1}{z}}$. Combining all these relations, we achieve the oscillation number as 
\begin{equation}
    \mathcal{N}=\frac{1}{z}\sqrt{\mathcal{E}-\mathcal{J}^2}-\mathcal{J}\tan^{-1}\left(\frac{\sqrt{\mathcal{E}-\mathcal{J}^2}}{\mathcal{J}}\right)
\end{equation}Expanding $\tan^{-1}\left(\frac{\sqrt{\mathcal{E}-\mathcal{J}^2}}{\mathcal{J}}\right)$ in powers of $\mathcal{J}$ and $\mathcal{E}$ and considering the large $\mathcal{E}$ limit we derive 
\begin{equation}
    \mathcal{N}=\frac{\sqrt{\mathcal{E}}}{z}-\frac{\pi\mathcal{J}}{z}+\left(1-\frac{1}{2z}\right)\frac{\mathcal{J}^2}{\sqrt{\mathcal{E}}}
\end{equation}This relation can further be inverted using the limit of large adiabatic oscillation number to generate the energy-angular momentum scaling for pulsating string ansatz as 
\begin{equation}
\begin{split}
    \mathcal{E}=&\mathcal{N}^2z^2+\frac{\mathcal{J}^2\pi^2z^2}{4}+\pi\mathcal{N}z^2\mathcal{J}+\left[\frac{\pi^2z^2}{4}+(1-2z)\right]\mathcal{J}^2\\&
    =\mathcal{N}^2z^2\left[1+\frac{\pi\mathcal{J}}{\mathcal{N}}+\frac{\mathcal{J}^2}{\mathcal{N}^2}\left(\frac{\pi^2}{2}-\frac{2}{z}+\frac{1}{z^2}\right)\right]
    \label{dispersion}
    \end{split}
\end{equation} Similarly as in rotating string ansatz, here also the scaling dispersion relation obtained by using the NR structure involves the critical exponent $z$ nontrivially.  
One may vary the energy with $z$, $\mathcal{N}$ and $\mathcal{J}$ individually by keeping other two parameters fixed at some chosen values. For all three cases, energy can be found to be monotonically increasing. 
For $z\rightarrow 1$, this reduces to 
\begin{equation}
     \mathcal{E}=\mathcal{N}^2+3.142\mathcal{N}\mathcal{J}+3.934\mathcal{J}^2
   \end{equation}where the leading order term goes like 
\begin{equation}
    \mathcal{E}\propto \mathcal{N}^2.
\end{equation}This scaling (\ref{dispersion}) is found to be consistent, at least upto the leading order, with that obtained for intrinsically nonrelativistic string pulsating in relativistic conformal geometry\cite{Roychowdhury:2019olt}. 
\section{Spin chain interpretation}
\label{section 4}
It is intriguing to learn that the energy spectra of both the rotating and pulsating string states  somewhat resemble those of some well-known frustrated and frustration-free spin chains which represent typical strongly correlated condensed matter systems. While frustrated spin chains consist of ground state degeneracy, it is removed in the frustration-free spin chains by introducing a boundary Hamiltonian along with the bare Hamiltonian. We will show in this section that, with the presence of anisotropy in the chosen Lifshitz target space, the rotating string ansatz retrieves a spin chain description of the frustrated spin-$\frac{1}{2}$ Heisenberg spin chain whereas the pulsating string states assume consistent interpretation from the perspective of frustration-free Fredkin or Motzkin spin chains.\subsection{Frustrated spin-$\frac{1}{2}$ $J_1-J_2$ Heisenberg chain from rotating string states}
 Energy in the scaling relation (\ref{scaling}) depends on $n,~~n\in\mathbb{Z}_+$. So the energy has discrete values for different $n$ when we fix $z$, $\omega$ and $J$. It is moreover straightforward that, keeping $J$ and $z$ fixed, there is a possibility of having degeneracy of energy states for the strings in the Lifshitz spacetime for some specific choices of $n$, say $n_1$ and $n_2$. This is expected to occur when the energy levels $n_1$ and $n_2$ satisfy the relation with frequencies $\omega_1$ and $\omega_2$ given as
\begin{equation}
\begin{split}
\sqrt{\omega_1^2z+n_1}=\sqrt{\omega_2^2z+n_2}
\end{split}
\label{degeneracy}
\end{equation}where, $n_1, n_2\in \mathbb{Z}_+$ and $n_1\neq n_2$. 
Such scenario of highly degenerate ground state  energy is similar to energy spectra of the frustrated $J_1-J_2$ spin chain model with ground state degeneracy\cite{OKAMOTO1992433,PhysRevB.51.11609}. $J_1-J_2$ spin chain is a generalization of the usual Heisenberg spin chain with nearest neighbour interaction and it contains an additional next-nearest-neighbour interaction with coupling not equal to that of the nearest-neighbour interaction. Similarly as the nearest-neigbour Heisenberg spin chain construction, the generalised frustrated spin-$\frac{1}{2}$ $J_1-J_2$ spin chain has been studied to be integrable with exact solutions in a series of literature \cite{HFrahm_1992,HolgerFrahm_1997,99a87bda6761453aa5f70376964c49d6,PhysRevLett.63.2524,PhysRevLett.59.2095,EKSklyanin_1988,Qiao:2019vsi}. For a simpler analysis, we consider the Majumdar-Ghosh model which is the simplest prototypical candidate in the family of frustrated $J_1-J_2$ spin chain. It describes 1D Heisenberg spin chain in which the nearest neighbor antiferromagnetic exchange interaction strength $J_1$ is twice as strong as the next-nearest neighbour interaction $J_2$\cite{WJCaspers_1984}, i.e.,  a spin-$\frac{1}{2}$ $J_1-J_2$ model with $J_1=2J_2$. At the point $J_1=2J_2$, such a Heisenberg spin chain has a frustrated disorder. At this point, it has a spin chain Hamiltonian as
\begin{equation}
    \mathcal{H}_{J_1-J_2}=\sum_i\frac{J_1}{2}\left(\vec{S}_i.\vec{S}_{i+1}+\vec{S}_{i}.\vec{S}_{i+2}+\vec{S}_{i+1}.\vec{S}_{i+2}\right),
\end{equation}$i$'s are summed over the sites on the spin chain. The corresponding ground state energy for both even and odd $i$ is given by 
\begin{equation}
    E_0=-\frac{3}{4}\frac{J_1}{2}
    \label{ground}
\end{equation}whereas, the other excited states have the energies 
\begin{equation}
    E_{\text{odd}}=-\frac{3}{4}\frac{NJ_1}{2},
    \label{excited}
\end{equation}$N$ being the number of sites on the spin chain. Eventually the ground state has two-fold degeneracy. The energy dispersion relation (\ref{scaling}) obtained from the construction of our desired integrable model assumes a similar form as (\ref{excited}) if we consider $J\equiv J_1$. In that case, the number of sites may be expressed by 
\begin{equation}
    N\approx \frac{16}{3\omega}\sqrt{\omega^2z+n},
\end{equation}provided that, for any fixed $z,\omega$ and $n$ we must take only the rounded integer value of the expression in right hand side as it represents the number of sites in the spin chain. 

Therefore, the solutions of NR integrable model emerging from rotating strings embeddings in a class of Lifshiz backgrounds yield the energy dispersion relation (\ref{scaling}) which shows degeneracy similar to the ground states of the frustrated spin-$\frac{1}{2}$ $J_1-J_2$ chain at its frustration points, while subjected to proper conditions. The integrability of such spin chain adds substantial support to the construction of integrable Neumann-Rosochatius model that generates such energy-momenta dispersion relations via its solutions. Following the similar procedure, we may have different sets of $\omega$- values for different $(n_1,n_2)$ to achieve high degeneracy of ground states of other suitable frustrated spin chains in the $J_1-J_2$ spin chain category. Nevertheless, in the $z\rightarrow 1$ limit, i.e., in the relativistic limit, $E=\pm \frac{2nJ}{\omega}$. This, for any specified $n$, reduces exactly to the form of the eigenvalues of the vacuum state of ferromagnetic Heisenberg spin chain. It is consistent with what one must expect from the energy-angular momentum dispersion relation of large $J$ family of rotating string solutions for Anti de-Sitter background.
\subsection{ Motzkin spin chain from pulsating string states}
The interpretation of energy as a function of oscillation number is not quite clear from the perspective of frustrated spin chain. Here we present a persuasive spin chain interpretation of such energy dispersion relation using the notion of frustration-free gapless spin chains. Fredkin and Motzkin spin chains are the classes of spin-1 and spin-$\frac{1}{2}$ gapless quantum spin chains respectively with frustration-free local nearest neighbor interaction Hamiltionians. These spin chains are studied in the context of the quantum Lifshitz model with $z=2$ as their ground states admit a direct correlation with the positive continuum Rokhsar-Kivelson states \cite{PhysRevLett.61.2376}, the later representing the ground states of free massless scalar Lifshitz field theories with any arbitrary values of $z$. 
    \label{fredkin}
Among these, the Motzkin spin chain is spin-$\frac{1}{2}$ frustration-free model that contains spin exchange with bilinear and biquadratic interactions \cite{Dell_Anna_2016}.
Its ground state is the superposition of the so-called Motzkin path Hamiltonian and it acquires the form
\begin{equation}
    H=\sum_{j=1}^{L-1}\Pi_{j,j+1}+H_{\text{bdy}}\,,
    \label{motzkin}
\end{equation}$\Pi_{j,j+1}$ describes the nearest neighbour interaction by incorporating different projectors. The Hamiltonian (\ref{motzkin}) consists of $U(1)$ symmetry. The boundary term plays a crucial role in choosing a unique ground state among the highly degenerate ground state manifold. After removing one of the local equivalence moves of the Motzkin paths, Motzkin spin chain acquires quantum integrability with specific boundary conditions \cite{Hao:2022juu}. This special type of Motzkin spin chain is coined as periodic free Motzkin spin chain. For Motzkin spin chain, there appears continuous parametric variation in the critical exponent $z$ for different low-lying energy excitations via DMRG analysis though the Hamiltonian remains gapless\cite{Chen_2017,Chen:2017txi}. From the explicit construction of Motzkin spin chain Hamiltonian in terms of spin operators we get eigenvalues of any chosen eigenstates. If we infer a match of these eigenvalues with the energy of the pulsating string states it will presumably relate the dynamical exponent $z$ with the large oscillation number $\mathcal{N}$ of the pulsating string. If we consider the eigenstate as a height variable $|h\rangle$ and operate different combinations of the spin operators in the basis $S^z$ then, following the explicit derivation of the interaction Hamiltonian in \cite{Chen_2017} 
and \cite{Chen:2017txi}, 
we can get constant eigenvalues for these states. Matching of these results with the energies of pulsating string states for different $z$ is expected to generate an inverse proportinality of $z$ to $\mathcal{N}$. This implies that, for large adiabatic oscillation number $\mathcal{N}$, the values of $z$ are small and also the energy gap tends to zero. This quite agrees with the scenario of low-lying excitations of gapless Motzkin spin chains. Along with the case of rotating string ansatz, plausible connections of the dispersion relation of the pulsating string solutions obtained from Neumann-Rosochatius framework with those of the integrable Motzkin spin chain can provide a good consistency check to remark about the integrability of our string-sigma model. A systematic study of the above spin chain interpretation of pulsating string can be a fascinating task for exploring their exact equivalence.  
\section{Summary and Conclusion}
\label{section 5}
We construct a 1D Neumann-Rosochatius-like  model by probing a regular fundamental string in a Poincare-type patch of nonrelativistic planar Lifshitz spacetime. We employ certain choices of embeddings for both rotating and pulsating  strings in Lifshitz spacetime so that the radial coordinate $r$ appears to be a function of the anisotropy index $z$. Despite the finite arbitrary anisotropy along the time direction, the model preserves an exactly solvable outlook that contains both harmonic oscillator-type and inverse square-type potentials in the associated Lagrangian and Hamiltonian with our chosen embeddings. However, the arbitrary critical exponent $z$ in the target space causes the harmonic oscillator in the solvable model to be constrained to move on a hyperbola whose eccentricity depends nontrivially on $z$. Moreover, we present the structures of the associated Uhlenbeck integrals of motion for our prescribed models constructed for strings both rotating and pulsating in the Lifshitz spacetime. It is oberserved that, due to finite anistotropy and therefore a constrained motion in a hyperbolic geometry with $z$-dependent eccentricity, the integrable framework can only be achieved if $z$ satisfies definite conditions involving the parameters $\kappa$, $T$, $m$ and the Lagrange multiplier $\Lambda_1$ for both rotating and pulsating string cases. Our exact solutions are then suitably utilized to derive the energy-momentum dispersion relations for both the chosen ansatz. The energy for rotating string states scales linearly with the momentum and the scale factor happens to be a nontrivial function of the critical exponent $z$. With $z\rightarrow 1$, i.e., in the limit where the conformal symmetry of the background is restored, the scaling relation reduces to that of the ferromagnetic Heisenberg spin chain which is the integrable toy model associated with the usual AdS/CFT holography. Furthermore, we comment on the behaviour of the energy with the variation of anisotropy in the target space background for both the chosen string ansatz. For the pulsating string, energy is expressed in terms of the power series of both momentum and the adiabatic invariant large oscillation number. Here also, there appears nontrivial presence of $z$ in the scaling expression of the energy. Finally we conjecture about some spin chain interpretations that can be straightforwardly speculated from the scalings of energy. For rotating case, the discreteness of the energy postulates its correlation with the heavily degenerate frustrated spin chains. In particular, we found similarities of our result with the energy of the degenerate ground states by considering the example of frustrated $J_1-J_2$ spin system with $J_2=\frac{J_1}{2}$. On the contrary, we discuss the energy of pulsating string states from the point of view of gapless Motzkin spin chains whose groundstates have direct correlations with the groundstates of Lifshitz field theories. Correlation functions of suitable Lifshitz field theory operators may be studied using the field-theoretic approach of spin-spin correlations in frustrated as well as frustration-free spin systems. From such point of view, these correlators of LFT are expected to be algebraic with an exponential decay at a long distance. The presence of such correlators indicates that the expected spin chain configuration from our computation is not highly disordered from its original Heisenberg structure. This implies that it must have the maximal number of individual spin flips. In \cite{Balents_2016}, 
authors proposed a universal field theoretic prescription of 1D frustrated ferromagnets by using the nonlinear sigma model technique in the vicinity of the quantum Lifshitz point. It would be rather interesting to study the algebraic correlators of the primary operators and the their scaling dimensions in such a field theory to check any possible one-to-one correspondence between those operators and different string states developed using the Neumann-Rosochatius integrable model. This will help to extract a dual picture of such string solutions with some plausible prototypical frustrated quantum magnets in the backdrop of the duality between the class of Lifshitz field theories and strongly correlated condensed matter systems. We would like to come back with some progress in this intriguing direction in near future.
\acknowledgments
AC would like to thank the  Research project supported by program "Excellence initiative – research university" for the AGH University for providing funds to carry out the above work. Ac acknowledges the insightful discussions with Dr. Aritra Banerjee and Nibedita Padhi during one of their recent work on nonrelativistic strings and exact solvability(2504.20252). These discussions immensely helped to build better understanding of the present work.
\bibliographystyle{JHEP}
\bibliography{NRLif}

    

\end{document}